\documentclass[12pt]{article}
\usepackage{amsmath}
\usepackage{cite}
\begin{document}
%%%%%%%%%%%%%%%%%%%%%%%%%%%%%%%%%%%%%%%%%%%%%%%%%%%
\def\thefootnote{\alph{footnote}}
\begin{flushright}
KANAZAWA-09-09  \\
April, 2010
\end{flushright}
\vspace{ .7cm}
\vspace*{2cm}
\begin{center}
{\LARGE\bf Neutrino masses and $\mu$ terms in a supersymmetric
extra U(1) model}\\
\vspace{1 cm}
{\Large Daijiro Suematsu}\footnote{e-mail:~suematsu@hep.s.kanazawa-u.ac.jp},
{\Large Takashi Toma}\footnote{e-mail:~t-toma@hep.s.kanazawa-u.ac.jp} 
{\Large and Tetsuro Yoshida}\footnote{e-mail:~yoshida@hep.s.kanazawa-u.ac.jp}
\vspace*{1cm}\\
{\itshape Institute for Theoretical Physics, Kanazawa University,\\
        Kanazawa 920-1192, Japan}\\
\end{center}
\vspace*{1cm}
{\Large\bf Abstract}\\
%%%%%%%%%%%%%%%%%Abstract%%%%%%%%%%%%%%%%%%%%%%%%%%%
We propose a supersymmetric extra U(1) model,
which can generate small neutrino masses and necessary $\mu$ terms,
simultaneously. Fields including quarks and leptons are embedded 
in three ${\bf 27}$s of $E_6$ in a different way among generations. 
The model has an extra U(1) gauge symmetry at TeV regions, 
which has discriminating features from other models studied previously. 
Since a neutrino mass matrix induced in the model has a constrained
texture with limited parameters, it can give a prediction. 
If we impose neutrino oscillation data to fix those parameters,
a value of $\sin\theta_{13}$ can be determined. 
We also discuss several phenomenological features which are 
discriminated from the ones of the MSSM.
%%%%%%%%%%%%%%%%%%%%%%%%%%%%%
\newpage
\section{Introduction} 
Recent experimental studies on neutrinos, the cosmic microwave 
background and the large scale structure of the universe have suggested 
the existence of neutrino masses \cite{oscil1, oscil2, oscil3, oscil4}
and dark matter \cite{dm1, dm2}. 
These give us strong motivation to
examine various possibilities for the extension of the standard model (SM).
If new neutral fields with suitable interactions are added to the SM, 
small neutrino masses can be generated and the origin of dark matter may
also be explained, simultaneously. In fact, the seesaw mechanism 
has been well known as such a typical example \cite{seesaw1, seesaw2, seesaw3}\footnote{In 
the $\nu$MSM proposed in \cite{nsm1, nsm2} 
which follows ordinary seesaw, since right-handed neutrinos are assumed to 
be lighter than the weak scale, the lightest right-handed neutrino can
behave as dark matter.} and 
also the radiative seesaw mechanism recently attracts much attention 
\cite{ma06}.
In the latter scenario, 
especially, new neutral fields added to
generate neutrino masses can behave as cold dark matter 
as long as it is the lightest stable field among newly introduced ones
\cite{ma06, scdm1, scdm2, scdm3, cdmmeg, fcdm1, fcdm2, fcdm3, ncdm1,
ncdm2, ncdm3, ext1, ext2, ext3, ext4, sty1, sty2}. However, the gauge hierarchy 
problem is put aside in the most study of these models.

On the other hand, in the minimal supersymmetric SM (MSSM) which is motivated 
to solve the gauge hierarchy problem, a dark matter candidate 
is automatically built 
in the model as the lightest neutralino as long as $R$-parity is assumed 
to be conserved \cite{susydm}.
However, even if singlet chiral superfields are
added as right-handed neutrinos in this model, 
small neutrino masses may not be explained without
assuming the existence of an intermediate scale of $O(10^{11-13})$ GeV 
as the origin of
the right-handed neutrino mass.\footnote{If we consider the
supersymmetric extension of the radiative seesaw model \cite{fks}, we
can have the right-handed neutrinos with $O(1)$~TeV masses.} 
Since the origin of that scale is not explained without 
further extension of the model, 
we have to consider other additional fields as in the grand unified models.
It is an interesting subject to find such a consistent extension  
without inducing other phenomenological problems. 
Proton stability, doublet-triplet splitting and also the $\mu$ problem 
are important issues to be answered in that
consideration.\footnote{The relation between the small neutrino masses
and the $\mu$ problem has been discussed in \cite{sy95, ds96, munu1,
munu2, munu3}, 
for example.}

In this paper we study these subjects from a view point of 
the neutrino mass generation 
in the framework of a supersymmetric extra U(1) model. 
In general, models with extra U(1) gauge symmetry are annoyed 
with gauge anomaly problem. However,
if we identify this U(1) symmetry with the ones derived from 
$E_6$ \cite{e6}, the 
anomaly problem can be solved automatically. It occurs as long as
the full members of the irreducible representation of $E_6$ 
are contained in the model.  
It is also well known that such supersymmetric models can appear
as the low energy effective models of heterotic string with 
Wilson line breaking \cite{w85,dkmns85}. 
An interesting point of these models is that the required low 
energy gauge symmetry can be derived from $E_6$ without any superfields of 
higher dimensional 
representations. Thus, the models can be constructed by strictly 
restricted fields only.

This type of model has a lot of interesting features.
One of them is that the models can have two U(1) gauge symmetries 
in addition to the U(1)$_Y$ in the SM \cite{dkmns85, ms861, ms862}. 
They can play important roles phenomenologically
at TeV scales or at intermediate scales of $O(10^{11-13})$ GeV.
For example, the $\mu$ problem has been shown to be solved elegantly 
by using this TeV scale extra U(1) symmetry \cite{sy95,cdeel}.
On the other hand, the intermediate scale can be introduced as the
breaking scale of another extra U(1) symmetry through a $D$-flat
direction. 
This scale can be related to the masses of the right-handed 
neutrinos \cite{sy95}.
It can also make extra matter fields heavy enough to decouple 
from low energy phenomena \cite{ds96, ms861, ms862}. 
However, if we set the model so as to use this intermediate scale 
for the mass generation of the right-handed neutrinos, 
we can not make extra color triplet fields and extra Higgs doublet 
fields heavy enough.
The extra U(1) symmetry constrained as a subgroup of $E_6$ forbids 
interaction terms required to make such fields heavy. 
Since these extra fields couple with quarks and leptons, 
dangerous couplings which induce proton decay and flavor changing
neutral current (FCNC) remains in the low energy regions 
in general.\footnote{In spite of this fault, 
this type of models have various interesting features.
Study on those points can be found in
\cite{u1ph-g1, u1ph-g2, u1ph-g3, u1ph-g4, u1ph-g5, u1ph-g6, u1ph-g7,
u1ph-g8, u1ph-g9, u1ph-g10, u1ph-g11, u1ph-h1, u1ph-h2, u1ph-h3,
u1ph-h4, u1ph-o1, u1ph-o2, u1ph-o3, u1ph-o4, u1ph-o5, u1ph-o6, u1ph-o7,
u1ph-o8, u1ph-o9, u1ph-o10, u1ph-o11, u1ph-o12, u1ph-o13, u1ph-c1, u1ph-c2, u1ph-c3, u1ph-c4}, for
example.}

In order to improve the difficulty of this type of extra U(1) model, 
we propose a novel modification.
In the models considered previously, the fields are assigned to a fundamental 
representation ${\bf 27}$ of $E_6$ in the same way
among the generation.
In this paper, we modify it among the generation
\cite{ds96} by imposing discrete symmetry on the superpotential of the model.
Under this setting, we show that this new model can solve the 
above mentioned various problems in the ordinary model of this type, 
simultaneously.\footnote{Only a type of extra U(1) at low energy 
regions has been known to generate small neutrino masses due to 
a high-scale seesaw mechanism in the $E_6$ framework.
It corresponds to a case with $\theta=\arctan\sqrt{15}$ 
if we define this extra U(1)$^\prime$
as U(1)$^\prime$=U(1)$_\chi\cos\theta$+U(1)$_\psi\sin\theta$. 
We note that our U(1) symmetry discussed in this paper corresponds 
to U(1)$_\chi$, which gives a new possibility for neutrino mass generation.} 
On the basis of this model, we derive a neutrino mass matrix and 
analyze it by using the neutrino
oscillation data. Other phenomenological features which are 
discriminated from the ones of the MSSM are also discussed.  

The remaining parts are organized as follows. In section 2 we define 
our model and discuss the discrete symmetry considered in the model.
We also address intermediate scales and explain its possible role in the model.
In section 3 we fix the effective model at the TeV regions and study
its various phenomenological features.
Neutrino mass generation in the model and results derived from it are
investigated in detail. In section 4 we summarize the paper.
 
\section{A model based on $E_6$}
\subsection{Field contents and symmetry}
We consider a model which is expected to be derived as an effective
model of string inspired $E_6$ models through Wilson line 
breaking \cite{w85,dkmns85}.
Gauge symmetry is SU(3)$\times$SU(2)$\times$U(1)$^3$ 
 which is expected to be induced from $E_6$ due to the Wilson line breaking.
Massless chiral superfields are composed of three of
the $E_6$ fundamental representation {\bf 27} and a part of the vector 
like pair ${\bf 27}+ \overline{\bf 27}$ \cite{dkmns85, ms861, ms862}. 
A fundamental representation {\bf 27} of $E_6$ can be decomposed 
under the above mentioned low energy gauge symmetry as shown in Table 1. 
As a remaining content of the vector like pair 
${\bf 27}+ \overline{\bf 27}$, we take 
${\cal A}_7+\bar {\cal A}_7$.\footnote{This type of field content has been
suggested to be induced through Wilson line breaking \cite{ms861, ms862}.}
Features of the model crucially depend on the way to embed the
physical fields in $A_1$ - $A_{11}$ of ${\bf 27}$. 
Although this embedding is usually done in the same 
way for every generation, we adopt here a twisted field assignment 
as shown in Table 1, where the fields are embedded differently 
among generations \cite{ds96}.

Gauge invariant renormalizable superpotential is given by
\begin{eqnarray}
W_1&=&A_1A_2A_8+A_1A_3A_9+A_4A_5A_9+A_4A_6A_8+A_7A_8A_9+A_7A_{10}A_{11} 
\nonumber \\
&+&A_1A_1A_{10}+A_1A_4A_{11}+A_2A_3A_{11}+A_2A_5A_{10}
+A_3A_6A_{10}\nonumber \\
&+&{\cal A}_7A_8A_9 + {\cal A}_7A_{10}A_{11},
\label{genw}
\end{eqnarray}
where Yukawa couplings and generation indices are abbreviated.
These terms generally exist unless some additional symmetry forbids them.
Since the terms in the second line are dangerous for 
proton stability, a part of those terms is required to disappear from
the low energy effective superpotential 
by imposing some additional symmetry or by 
making extra fields sufficiently heavy 
as a result of symmetry breaking at some high energy scales.
As easily seen from Table 1, the model has three color triplet 
pairs $(g, \bar g)$
and two extra Higgs doublet pairs $(H_u, H_d)$ beyond the MSSM contents
at this stage.

It is important to note that there also appear gauge 
invariant nonrenormalizable terms
\begin{equation}
W_1^{NR}=\sum_{(a,b,c)}\left(\frac{{\cal A}_7\bar{\cal A}_7}{M_{\rm pl}^2}
\right)^{n_{abc}}A_aA_bA_c\equiv
\sum_{(a,b,c)}\epsilon^{n_{abc}}A_aA_bA_c,
\qquad (A_aA_bA_c \in W_1),
\label{nr}
\end{equation}
where $M_{\rm pl}$ is the Planck mass and $n_{abc}$ is a positive
integer to be determined for each term $A_aA_bA_c$ in $W_1$
independently.
The value of $n_{abc}$ depends on the additional symmetry considered 
in the model.
Some of these can cause important contributions to the 
low energy effective models
if both ${\cal A}_7$ and $\bar{\cal A}_7$ obtain large vacuum
expectation values (VEVs). 
In particular, the terms in $W_1^{NR}$ corresponding to the last two 
terms in $W_1$ should be taken into 
account since they are relevant to the mass of the above mentioned extra
fields.
These points will be discussed in the next subsection. 

\begin{figure}[t]
\begin{center}
\begin{tabular}{|c|c|c|c|c|}\hline
    & SU(3)$\times$SU(2)$\times$U(1)$^3$ 
& \multicolumn{2}{c|}{Field
 assignment($Z_2\times Z_4$)}& Light fields\\
\hline\hline
$A_1$& $({\bf 3},{\bf 2},
\frac{1}{6}, 1, -1)$  
& \hspace*{3mm}  
$Q_\alpha~(-,+1)$\hspace*{3mm}  & $Q_3~(-,+1)$  &  $Q_i$ \\
$A_2$ & $({\bf 3}^\ast,{\bf 1},
-\frac{2}{3}, 1, -1)$ 
& $\bar U_\alpha~(-,+2)$ & $\bar U_3~(-,+2)$ 
& $\bar U_i$ \\
$A_3$ & $({\bf 3}^\ast,{\bf 1},
\frac{1}{3}, 1, 3)$ 
& $\bar D_\alpha(-,0)$ & $\bar D_3~(-,0)$ & 
$\bar D_i$\\
$A_4$ & $({\bf 1},{\bf 2},
-\frac{1}{2}, 1, 3)$ 
& $L_\alpha~(-,+1)$ & $H_{d3}~(+,+2)$ & 
$L_\alpha,~H_{d3}$ \\
$A_5$ & $({\bf 1},{\bf 1},
1, 1, -1)$ 
& $\bar E_\alpha~(-,0)$ & $\bar E_3~(-,0)$ & 
$\bar E_i$\\
$A_6$ & $({\bf 1},{\bf 1},
0, 1, -5)$ 
& $\bar N_{\genfrac{}{}{0pt}{2}{1}{2}}~(+,\genfrac{}{}{0pt}{2}{0}{+1})$ 
& $S_3~(+,+1)$ & 
$\bar N_\alpha,~S_3$\\
$A_7$ & $({\bf 1},{\bf 1},
0, 4, 0)$ 
& $S_{\genfrac{}{}{0pt}{2}{1}{2}}~(+,\genfrac{}{}{0pt}{2}{+2}{0})$ 
& $\bar N_3~(-,+2)$ & 
$S_2$\\
$A_8$ & $({\bf 1},{\bf 2},
-\frac{1}{2}, -2, 2)$ 
& $H_{u_{\genfrac{}{}{0pt}{2}{1}{2}}}~(+,\genfrac{}{}{0pt}{2}{+2}{+1})$ 
& $H_{u3}~(-,+2)$ & 
$H_{u\alpha}$ \\
$A_9$  & $({\bf 1},{\bf 2},
-\frac{1}{2}, -2, -2)$ 
& $H_{d_{\genfrac{}{}{0pt}{2}{1}{2}}}~(\genfrac{}{}{0pt}{2}{-}{+}
,\genfrac{}{}{0pt}{2}{+1}{+3})$ 
& $L_3~(-,+2)$  
& $H_{d2},~L_3$ \\
$A_{10}$  & $({\bf 3},{\bf 1},
-\frac{1}{3},- 2, 2)$ 
& $g_{\genfrac{}{}{0pt}{2}{1}{2}}~(\genfrac{}{}{0pt}{2}{-}{+},
\genfrac{}{}{0pt}{2}{+1}{+2})$ 
& $g_3~(-,+3)$ & $g_3$\\
$A_{11}$  & $({\bf 3}^\ast,{\bf 1},
\frac{1}{3}, -2, -2)$ 
& $\bar g_{\genfrac{}{}{0pt}{2}{1}{2}}~(\genfrac{}{}{0pt}{2}{-}{+},
\genfrac{}{}{0pt}{2}{+2}{+1})$ 
& $\bar g_3~(-,+1)$ 
& $\bar g_3$ \\ \hline
\end{tabular}
\end{center}
\vspace*{-1mm}
{\footnotesize Table~1 Decomposition of {\bf 27} and the field
 assignment. Abelian charges $Y$, $Q_\psi$ and $Q_\chi$ for
 U(1)$_Y\times$U(1)$_\psi\times$U(1)$_\chi$ are listed as
$\sqrt{\frac{5}{3}}Y, ~2\sqrt{10}Q_\psi$ and $2\sqrt{6}Q_\chi$, 
respectively \cite{e6}.
Greek and Latin indices of the fields stand for the generation.
On the discrete symmetry, we show the $Z_2$ parity and 
the charge for $Z_4$ of each field, respectively.}
\end{figure}

Now we consider the model with $Z_2\times Z_4$ as the 
additional symmetry.\footnote{
Since we suppose the Wilson line breaking of $E_6$, 
this discrete symmetry should be consistent with the gauge symmetry 
SU(3)$\times$SU(2)$\times$ U(1)$^3$ but not with $E_6$. 
We may be able to find relations between this
symmetry and the discrete symmetry which is used to define the multiply
connected manifold as the basis of 
Wilson line breaking as shown in \cite{ms861, ms862}. It is likely that they 
are identified each other and in that case the present discrete symmetry
is considered to be built in the model originally.}
Its charge assignment for each chiral superfield 
in ${\bf 27}$ are also shown in Table.~1.
The $Z_2$ symmetry will be identified with $R$ parity.
For ${\cal A}_7$ and $\bar {\cal A}_7$, 
we assign them the $Z_2$ parity and the $Z_4$ charge as $(+,+1)$ and $(+,0)$, 
respectively. 
The allowed terms in superpotential $W_1$ under this discrete symmetry 
can be restricted as follows,
\begin{eqnarray}
W_2&=&Q_i\bar U_j H_{u_2} +Q_i\bar D_j H_{d_2}+L_\alpha \bar E_j H_{d_2} 
+L_3 \bar E_j H_{d_3}
+ S_2 H_{u_2}H_{d_2} + S_3H_{u_2}H_{d_3}    \nonumber \\ 
&+& S_2 g_3\bar g_3 + \bar N_3 g_1\bar g_2 + Q_i\bar g_3 H_{d_3}
+ g_3\bar D_i S_3 + g_1\bar D_i S_3
+ g_3\bar D_i\bar N_2 \nonumber \\
&+&L_\alpha S_3 H_{u_3}+ L_3 S_2 H_{u_3}+L_\alpha\bar N_2 H_{u_3}
+\bar N_1 H_{u_1}H_{d_3}+\bar N_2 H_{u_2}H_{d_3}  \nonumber \\
&+& \bar N_3 H_{u_2}H_{d_1}
+{\cal A}_7H_{u3}H_{d_1}+ Q_iQ_jg_2 +\bar U_i\bar E_j g_2
+{\cal A}_7g_\alpha\bar g_\alpha,
\label{z2invw}
\end{eqnarray}
where Yukawa couplings are omitted again.
Generation indices are labeled 
by Greek and Latin characters $\alpha=1,2$ and $i=1,2,3$.
Nonrenormalizable terms allowed by the same symmetry in $W_1^{NR}$ 
are easily found to be restricted to
\begin{eqnarray}
W_2^{NR}&=&\epsilon^2{\cal A}_7H_{u_1}H_{d_2}
+\epsilon^3{\cal A}_7H_{u_2}H_{d_2}
+\epsilon^3{\cal A}_7L_3H_{u_3} 
+\epsilon^3{\cal A}_7g_3\bar g_{3}
+\epsilon{\cal A}_7g_1\bar g_{3}+\epsilon^2{\cal A}_7g_3\bar g_{1} \nonumber \\
&+&\epsilon\left( 
H_{d_3}\bar E_iH_{d_1}+L_\alpha\bar N_1H_{u_3}+H_{d_3}\bar N_1H_{u_2}
+S_1H_{u_1}H_{d_2}+S_2H_{u_3}H_{d_1} \right. \nonumber \\
&+&\left.\bar N_3H_{u_3}H_{d_2}+S_2g_\alpha\bar g_\alpha 
+Q_iL_\alpha\bar g_2 + \bar U_i\bar D_j\bar g_2+\bar D_i\bar N_1g_3\right)
+O(\epsilon^2).
\label{nr2}
\end{eqnarray}
In eq.~(\ref{nr2}), we list up all terms corresponding to the ones 
in $W_1$ which include ${\cal A}_7$. 
For other terms, we write only the $O(\epsilon)$ terms in eq.~(\ref{z2invw}).
The reason for this is that the VEV of ${\cal A}_7$ is considered large
enough but $\epsilon\ll 1$, as shown in the next subsection.

As found in these superpotentials $W_2$ and $W_2^{\rm NR}$, 
there still remain the couplings
among extra colored fields and quarks which are dangerous for proton stability.
Moreover, small neutrino masses can not be explained only from these.
The existence of extra fields could also spoil gauge coupling
unification. Because of these reasons, it seems to be favorable that 
a part of these extra fields becomes heavy through some symmetry 
breaking at a much higher energy scale than the weak scale. 
If the fields contributing to neutrino mass generation become heavy 
enough due to the same symmetry breaking, the smallness of neutrino masses 
can also be explained. 
In the next part, we address the possibility to cause this symmetry
breaking at a desirable high energy scale. 

\subsection{An intermediate scale induced by a $D$-flat direction}
Problems addressed at the end of the last part
can be solved if scalar components of a vector like pair 
${\cal A}_7+\bar{\cal A}_7$ have large VEVs. 
The lowest order invariant superpotential for ${\cal A}_7$ 
and $\bar{\cal A}_7$ is written as 
\begin{equation}
W_{\cal A}=\frac{c}{M_{\rm pl}^5}
\left({\cal A}_7\bar{\cal A}_7\right)^4,
\label{intw}
\end{equation}
where the coupling constant $c$ is naturally considered to be $O(1)$. 
Since $\langle {\cal A}_7\rangle=\langle \bar{\cal A}_7\rangle=\phi$ 
gives a $D$-flat direction of extra U(1) gauge symmetries, minimum
points of the scalar potential for these are expected to appear 
along this direction \cite{dkmns85, ms861, ms862}. 
On these points, both the extra gauge symmetry U(1)$_\chi$
and the $Z_2$ symmetry remain unbroken. 
Although $Z_4$ is spontaneously broken, an associated domain wall 
problem can not be
serious in this case. Since this symmetry breaking occurs at
sufficiently high energy scale as seen below, inflation is expected to
occur after it to resolve this problem.
The remaining U(1)$_\chi$ is expected to give a solution to 
the $\mu$ problem at TeV scales in the way as suggested in \cite{sy95}.
Moreover, since the $Z_2$ parity of the SM contents are assigned as
even, the lightest $Z_2$ odd field is stable to be a dark matter candidate. 
    
The scalar potential derived from $W_{\cal A}$ along this $D$-flat direction 
is expressed by the VEV $\phi$ as
\begin{equation}
V\simeq \frac{32}{M_{\rm pl}^{10}}|\phi|^{14}-2m_s^2|\phi|^2,
\end{equation}
where $m_s$ stands for the soft supersymmetry breaking mass of $O(1)$~TeV.
By minimizing the scalar potential $V$, we can determine a value of
$\phi$ as 
\begin{equation}
|\phi|\simeq \left(0.2 M_{\rm pl}^5m_s\right)^{\frac{1}{6}}.
\end{equation}
This gives $|\phi|\sim 2\times 10^{16}$~GeV and then $\epsilon$ defined in 
eq.~(\ref{nr}) is estimated as $\epsilon\sim 10^{-6}$. Since fermionic 
components of ${\cal A}_7$ and $\bar{\cal A}_7$ mix with 
a broken U(1)$_\psi$ gaugino, their mass eigenvalues are expected to be 
$O(g_\psi|\phi|)$ where $g_\psi$ is a gauge coupling constant of U(1)$_\psi$.
 
The symmetry breaking at this scale controls the massless field 
contents which constitute the low energy effective model through the couplings
with ${\cal A}_7$. This is found in $W_2$.
Although the model has three pairs of Higgs doublets originally,
only a part of them can remain massless. 
In fact, a gauge invariant coupling ${\cal A}_7H_{u_3}H_{d_1}$ 
can induce mass terms between Higgs chiral superfields $H_{u_3}$ 
and $H_{d_1}$.  
On the other hand, since ${\cal A}_7$ has also gauge invariant couplings with 
extra colored fields as ${\cal A}_7g_\alpha\bar g_\alpha$,\footnote{
This coupling of extra colored fields with ${\cal A}_7$ may be 
relevant to a solution for the strong CP problem \cite{strongcp1, strongcp2}. } 
the VEV $\phi$ generates large masses for them. Thus, only 
one pair of extra color triplets remains almost massless.   
We find that the mass of the states dominated by $g_3$ or 
$\bar g_3$ is $O(\epsilon^3|\phi|)$ by removing the mixings 
$\epsilon^2{\cal A}_7g_3\bar g_1$ and $\epsilon{\cal A}_7g_1\bar g_3 $ 
appearing in $W_2^{NR}$.
The VEV $\phi$ also makes some of singlets $\bar N_i$ and $S_i$ 
very heavy through another gauge invariant nonrenormalizable coupling
\begin{equation}
\frac{1}{M_{\rm pl}}(\bar{\cal A}_7A_7)^2.
\label{majmass}
\end{equation}
This coupling is controlled by the discrete symmetry $Z_4$ 
and generates the mass of $O(|\phi|^2/M_{\rm pl})$ for $\bar N_3$ and $S_1$, 
which may play a role of heavy right-handed neutrinos with the mass of
$O(10^{13})$ GeV \cite{sy95}.
 
Here, it is worthy to note the magnitude of the 
couplings with ${\cal A}_7$ in other terms, which are forbidden 
by $Z_4$ as the renormalizable couplings in $W_2$ 
but appear as the invariant 
nonrenormalizable couplings in $W_2^{NR}$. 
Since $\epsilon^3{\cal A}_7H_{u_2}H_{d_2}$ and 
$\epsilon^3{\cal A}_7H_{u_3}L_3$ are sufficiently suppressed,
they can be neglected in the following discussion.
On the other hand, since $\epsilon^2{\cal A}_7H_{u_1}H_{d_2}$ induces a
weak scale mass term for $H_{u_1}$ and $H_{d_2}$ if Yukawa coupling is a
little small, 
it can play an important role in the low energy phenomenology. 

Taking account of the facts discussed above, we see that the massless chiral 
superfields in the model are confined to the ones listed in the last 
column of Table~1.
At the scale of $\phi$, 
they are composed of the MSSM contents
and also the extra chiral superfields which are summarized as 
$({\bf 5},~\bar{\bf 5}) + 4({\bf 1})$ of SU(5). 
These massless contents keep the unification of the MSSM gauge 
coupling constants at $M_{\rm GUT}\simeq 2\times 10^{16}$~GeV,
since the addition of ${\bf 5}+\bar{\bf 5}$ to the MSSM changes the
one-loop $\beta$-function of each MSSM gauge coupling constant in the
same way.
It is noticeable that the symmetry breaking scale obtained above happens 
to coincide with the scale of this $M_{\rm GUT}$. 

\section{A low energy effective model} 
Phenomenological features of the low energy effective model obtained 
from the discussion
in the previous section are determined by the superpotential 
which is composed of the light chiral superfields,\footnote{In the following
parts, the scalar component is expressed by the same character as the
chiral superfield and a tilde is put on the character for the fermionic
component.}
\begin{eqnarray}
W_3&=&h_U^{ij}Q_i\bar U_j H_{u_2} +h_D^{ij}Q_i\bar D_j H_{d_2}
+h_E^{\alpha j}L_\alpha \bar E_j H_{d_2} 
+h_E^{3j}L_3 \bar E_j H_{d_3} \nonumber\\
&+&\lambda_1 S_2 H_{u_2}H_{d_2} +\lambda_2 S_3H_{u_2}H_{d_3} 
+k S_2 g_3\bar g_3 +h_{\bar g}^iQ_i\bar g_3 H_{d3}+h_g^ig_3\bar D_i S_3
 \nonumber \\
&+&f_\alpha L_\alpha S_3 H_{u3}+ f_3 L_3 S_2 H_{u_3}
+ f_\alpha^\prime L_\alpha\bar N_2 H_{u_3}  \nonumber \\
&+& \kappa_1 \bar N_1 H_{u_1}H_{d_3}
+\kappa_2\bar N_2 H_{u_2}H_{d_3}+\kappa_i^\prime g_3\bar D_i\bar N_1.
\label{lspot}
\end{eqnarray}
This superpotential $W_3$ includes necessary terms to realize the
favorable structure of the MSSM and
also the important terms for the neutrino mass generation.
It should be noted that the third line in $W_3$ contains the terms which 
include heavy chiral superfields relevant to the 
neutrino mass generation.
As soft supersymmetry breaking parameters, we introduce 
the soft masses $m_s^2$ for scalar components of all
chiral superfields, and a trilinear scalar coupling $A$ with mass dimension one 
which appears for each term in $W_3$, and the soft masses of gauginos 
$\tilde W_i,~\tilde B,~\tilde\chi$ for each gauge factor group SU(2), U(1)$_Y$,
U(1)$_\chi$ and also gluinos.
All other mass scales in the model are considered to be radiatively generated 
from these soft supersymmetry breaking parameters. 
Now, we address several interesting features of this model 
in the following parts.

\subsection{$\mu$ terms and quark/lepton masses}
The model has two kinds of $\mu$ term. 
Following the radiative symmetry breaking scenario based on
the renormalization group equations (RGEs) for the soft scalar masses 
of the singlet scalars $S_{2,3}$, 
they can obtain the VEVs as follows,
\begin{equation}
\langle S_2\rangle=u, \qquad \langle S_3\rangle=u^\prime.
\label{svev}
\end{equation}
This is expected to occur since these singlet scalars have couplings 
with the colored fields and then the squared masses of $S_{2,3}$ 
can take negative values at weak scales \cite{sy95,cdeel}. 
As a result of these VEVs, two $\mu$ terms are generated from 
the $\lambda_{1,2}$ terms as $\mu=\lambda_1 u$ 
and $\mu^\prime=\lambda_2 u^\prime$. 
Since $H_{u_2}$ has a large Yukawa coupling with top quark, 
it is also expected to have a VEV as usual. 
Thus, if we note the existence of effective $\mu$ terms and remind the
experience in the MSSM, the vacuum structure at TeV regions 
is considered to be fixed by 
\begin{equation}
\langle H_{u_2}\rangle =v_2,\quad \langle H_{u_3}\rangle =0,
\quad \langle H_{d_2}\rangle =v_{1a}, \quad
\langle H_{d_3}\rangle =v_{1b},
\label{vev}
\end{equation}
where we define $v_1^2=v_{1a}^2+v_{1b}^2$ and $\tan\beta=v_2/v_1$.
The $Z_2$ symmetry remains as an exact symmetry in this vacuum. 
It is identified with the $R$-parity,
since the charge can be assigned such that 
the SM contents are even and their superpartners are odd under it.
It is useful to note that the usual $R$-parity violating terms 
allowed by the MSSM gauge structure are also forbidden 
at the level of $W_2$ through the existence of the extra U(1) symmetries.
This $Z_2$ symmetry can guarantee the stability of the lightest particle
with odd parity.

Under this vacuum, the masses of quarks and charged leptons 
are generated as
\begin{equation}
(m_u)^{ij}=h_U^{ij}v_2, \quad (m_d)^{ij}=h_D^{ij}v_{1a}, \quad
(m_e)^{ij}=\left(\begin{array}{ccc}
h_E^{\alpha 1}v_{1a} & h_E^{\alpha 2}v_{1a} & h_E^{\alpha 3}v_{1a} \\
h_E^{31}v_{1b} & h_E^{32}v_{1b} & h_E^{33}v_{1b} \\
\end{array} \right).
\label{qlmass}
\end{equation}
Neutrino masses are discussed in the next part.
The charged lepton mass matrix is generated by the couplings with two Higgs
doublets $H_{d_{2,3}}$. This could induce dangerous lepton flavor
violating processes in principle.
However, it can be easily escaped as long as the Yukawa coupling
constants satisfy simple relations given in Appendix A.
This comes from a feature of the model that each lepton doublet 
originally couples with only one Higgs doublet. 
In fact, as shown in Appendix A, we can find the basis of right-handed 
charged leptons such that the matrix for the Yukawa couplings takes the 
block diagonal form 
\begin{equation}
\tilde h_E^{\alpha\beta}L_\alpha\bar E_\beta^\prime H_{d_2}
+\tilde h_E^{33}L_3\bar E_3^\prime H_{d_3}.
\label{diag}
\end{equation}
This new basis is consistent with the 
imposed discrete symmetry.
Thus, there appears no additional lepton flavor mixing induced through
the light Higgs sector under the supposed conditions.
  
The down quark sector has the mass mixing with the 
extra colored fields as
\begin{equation}
(\tilde Q_i~ \tilde g_3)\left(\begin{array}{cc} 
(m_d)^{ij} & h^i_{\bar g}v_{1b}\\
 h^j_g u^\prime & ku \\   \end{array}\right)
\left(\begin{array}{c}\tilde{\bar D}_j \\ \tilde{\bar g}_3\\ \end{array}\right).
\label{dmass}
\end{equation}
This is an important feature of this model. 
The mixing may be detected as the deviation from the CKM scheme 
through future experiments for the $B$ meson system.
These extra colored fields are also expected to be found at the LHC directly.
If each Yukawa coupling takes appropriate values, these mass matrices
(\ref{qlmass}) and (\ref{dmass}) are expected to 
realize the mass eigenvalues and the flavor mixings. 
However, since the model can not predict the magnitude and flavor structure of 
the Yukawa couplings at the present stage, we do not discuss this 
problem further here.

\subsection{Mass and mixing of neutrinos}
Next, we proceed to estimate the mass and the mixing of neutrinos in the model.
For this purpose, it is useful to investigate the nature of other 
neutral fermions in this effective model.
We have $Z_2$ even heavy neutral fermions $\tilde H_{u_3}$, $\tilde H_{d_1}$ and
$\tilde{\bar N}_3$ in addition to the ordinary left-handed neutrinos 
$\nu_i(\equiv \tilde{L}_i^0)$.
Their masses are induced through the VEV $\phi$. 
This means that their $Z_2$ odd scalar partners are sufficiently heavy. 
On the other hand, we have Higgsinos $\tilde H_{u_{1,2}},~\tilde H_{d_{2,3}}$, 
singlinos $\tilde S_{2,3},~\bar N_{1,2}$, and gauginos 
$\tilde W_3,~\tilde B,~\tilde\lambda_\chi$ as $Z_2$ odd neutral 
fermions which constitute a part of neutralinos. 
Since the scalar partners of the Higgsinos and the singlinos except 
for $H_{u_1}$ and $\bar N_{1,2}$ have the VEVs as discussed before, 
they can obtain the masses through the effective $\mu$ terms and mix 
with other $Z_2$ odd neutral fermions including gauginos.
The mass matrices of these neutral fermions are given in Appendix B. 
Since all of these neutralinos can have weak scale masses, 
the model can satisfy the experimental constraint imposed by the 
$Z^0$ invisible width. 
If we remind that this $Z_2$ symmetry remains as the exact one 
even after the symmetry 
breaking at the weak scale, we find that the left-handed neutrinos $\nu_i$ 
can not mix with these neutralinos. 
Thus, a dark matter candidate in this model is the lightest 
neutralino which can have different features from the one in the MSSM, in
principle.

\input epsf
\begin{figure}[t]
\begin{center}
\epsfxsize=14cm
\leavevmode
\epsfbox{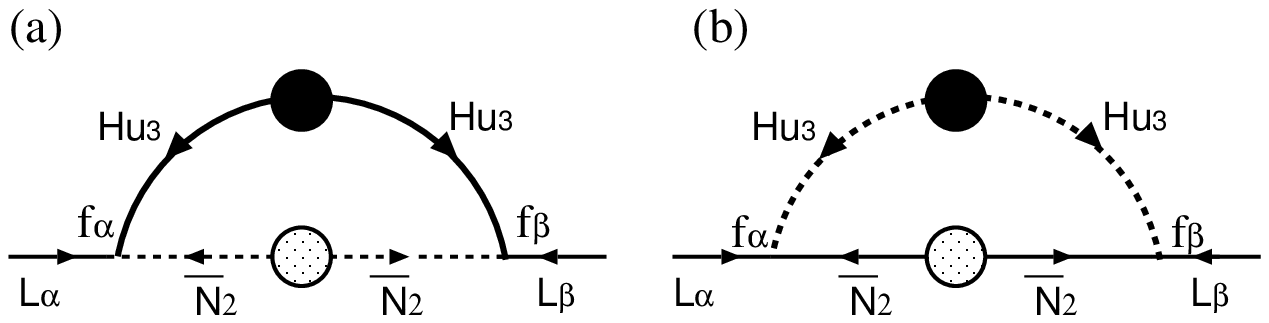}
\vspace*{-4mm}
\end{center}
{\footnotesize {\bf Fig.~1}~~Diagrams contributing to the neutrino
Majorana masses. the bulbs in internal fermion lines 
of (a) and (b) are induced 
through the neutral fermion mass matrices in 
eqs.~(\ref{h3}) and (\ref{neutram}). 
On the other hand, the bulbs in internal scalar lines of (a) and (b)
is induced by
supersymmetry breaking $A$ terms in eqs.(\ref{ah3}) and (\ref{bmass}), 
respectively. }
\end{figure}

The present model contains a heavy $Z_2$ even neutral fermion $\tilde H_{u_3}^0$
and its scalar partner which has no VEV. 
Here we note that Majorana mass is generated for $\tilde H_{u_3}^0$
through the mixing with the heavy fermion $\tilde{\bar N}_3$. 
Since they can couple with left-handed neutrinos $\nu_i$ 
as found in $W_3$, small neutrino masses and
non-trivial neutrino mixing can be induced.
If the singlet scalars $S_{2,3}$ obtain the VEVs as shown 
in eq.~(\ref{svev}), the first two terms in the third
line of $W_3$ generate Majorana masses for $\nu_{1,2,3}$ through 
the seesaw mechanism.
Unfortunately, only one nonzero mass eigenvalue can be generated
from these contributions.
However, we find that $\nu_{1,2}$ can also obtain the radiative mass 
through one-loop diagrams as shown in Fig.~1.
These effects make the model viable for the neutrino mass generation.  

The mass matrix for three light Majorana neutrinos $\nu_{1,2,3}$ 
is expressed by
\begin{equation}
M_\nu=\Lambda_1\left(\begin{array}{ccc}
y_1^2 & y_1y_2 & y_1\\
y_1y_2 & y_2^2 & y_2 \\ 
y_1 & y_2 & 1 \\ \end{array}\right)
+(\Lambda_2^a+\Lambda_2^b)\left(\begin{array}{ccc}
\bar f^{2} & \bar f & 0\\
\bar f  & 1 & 0\\
0 & 0 & 0\\ \end{array}\right),
\label{neutmass}
\end{equation} 
where $y_\alpha=f_\alpha u^\prime/f_3u$ and 
$\bar f=f_1^\prime/f_2^\prime$.
The first term is generated through the ordinary seesaw mechanism 
caused by the heavy neutral fermion $\tilde H_{u_3}^0$.
The second term is generated through the radiative seesaw mechanism and 
their relevant diagrams are shown in Fig.~1.\footnote{The similar type of 
radiative neutrino mass generation in supersymmetric model has been 
discussed in \cite{susyma1, susyma2}. 
However, details are different between the present model and the one 
discussed there. 
As mentioned in the previous parts, the present extra U(1)s make 
it possible to solve the $\mu$ problem radiatively and also give 
both the origins of the existence of the singlet chiral superfields and the 
right-handed neutrino mass scale.} 
The mass scales $\Lambda_1$  and $\Lambda_2^{a,b}$ are estimated as   
\begin{eqnarray}
&&\Lambda_1=\frac{(f_3u)^2}{M}, \qquad
\Lambda_2^a=\frac{(\Lambda_B^a)^2f_2^{\prime 2}}
{8\pi^2m_F^a}I\left(\frac{(m_F^a)^2}{(m_B^a)^2}\right), \qquad
\Lambda_2^b=\frac{(\Lambda_B^b)^2f_2^{\prime 2}}
{8\pi^2m_F^b}I\left(\frac{(m_F^b)^2}{(m_B^b)^2}\right),
 \nonumber \\
&&I(x)=\frac{x}{1-x}\left(1+\frac{x}{1-x}\ln x\right)
\label{scale}
\end{eqnarray} 
where $M$ is the effective mass of $H_{u_3}$, which is generated 
by the VEV $\phi$. In eq.~(\ref{scale}),
$m_B^a(m_B^b)$ and $\Lambda_B^a(\Lambda_B^b)$ stand for the mass and 
the mixing shown by the bulbs in the internal scalar lines in 
Fig.~1(a)(Fig.~1(b)), respectively. 
The effective mass of the internal fermion is expressed by $m_F^{a,b}$.
Detailed discussion on these issues is given in Appendix B. 
Since both the fermionic component $\tilde H_{u_3}$ and the bosonic component 
$H_{u_3}$ obtain the large masses due to  
the VEV $\phi$, $m_F^a \gg m_B^a$ is satisfied in Fig.~1(a) 
and also $m_B^b\gg m_F^b$ in Fig.~1(b).
Using this fact, $\Lambda_2^{a,b}$ can be estimated for each case as
\begin{eqnarray}
{\rm (a)}&&\Lambda_2^a\simeq \frac{(\Lambda_B^a)^2f_2^{\prime 2}}{8\pi^2m_F^a}
\ln\frac{(m_F^a)^2}{(m_B^a)^2}
\simeq \frac{f_2^{\prime 2}\kappa_2^2 A^3\mu^\prime v_2v_{1b}}
{8\pi^2 \gamma_1\phi m_s^4}
\ln\left(\frac{(\gamma_1\phi)^2}{m_s^2}\right), \nonumber \\
{\rm (b)}&&\Lambda_2^b\simeq \frac{(\Lambda_B^b)^2f_2^{\prime 2}m_F^b}
{8\pi^2(m_B^b)^2}
\simeq\frac{f_2^{\prime 2}A^5 (\gamma_2v_2)^2m_w}{8\pi^2 (\gamma_1\phi)^4M_N^3}.
\end{eqnarray}
From these expressions, we find that the diagram (a) 
gives the dominant contribution
to the neutrino masses. This scale can be desirable values if
the couplings and the soft supersymmetry breaking parameters 
are fixed appropriately within the reasonable regions.
Since the VEV $\phi$ is sufficiently large as discussed before, 
we do not need any unnaturally 
small coupling constants to generate the small neutrino masses. 
In this point the present model improves the original non-supersymmetric 
version for the radiative seesaw model \cite{ma06}.

\begin{figure}[t]
\begin{center}
\epsfxsize=8cm
\leavevmode
\epsfbox{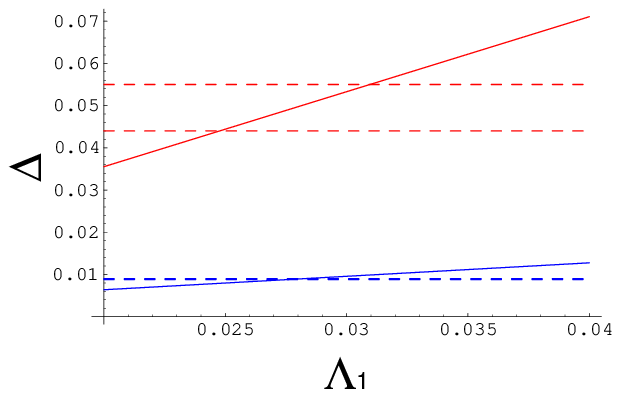}
\vspace*{-5mm}
\end{center}
{\footnotesize {\bf Fig.~2}~~Mass eigenvalues as functions of
 $\Lambda_1$ in the normal hierarchy case 
with $\sin\theta_{13}=0.13$. 
Red and blue solid lines represent $\Delta_{31}\equiv\sqrt{m_3^2-m_1^2}$
and $\Delta_{21}\equiv\sqrt{m_2^2-m_1^2}$, respectively. 
The regions of $\sqrt{\Delta m_{\rm atm}^2}$ and 
$\sqrt{\Delta m_{\rm sol}^2}$ required by the neutrino oscillation data are 
plotted by the red and blue dashed lines. All mass units are taken as eV.}
\end{figure}
 
We note that the mass matrix $M_\nu$ is expressed by using only 5 
independent parameters. This makes the model predictive. 
If we impose the known experimental data, we can predict 
a value of $\sin\theta_{13}$.     
We examine the validity of this neutrino mass matrix by imposing the
neutrino oscillation data. 
We know how neutrinos should mix each other by using the neutrino 
oscillation data \cite{oscil1, oscil2, oscil3, oscil4}. We require that the neutrino 
mass matrix $M_\nu$ is diagonalized by
\begin{equation}
U=\left(\begin{array}{ccc}
1 & 0 & 0 \\
0 & \cos\theta_{23} & \sin\theta_{23} \\
0 & -\sin\theta_{23} & \cos\theta_{23} \\
\end{array}\right)
\left(\begin{array}{ccc}
\cos\theta_{13} & 0 & \sin\theta_{13} \\
0 & 1 & 0 \\
-\sin\theta_{13} & 0 & \cos\theta_{13} \\
\end{array}\right)
\left(\begin{array}{ccc}
\cos\theta_{12} & \sin\theta_{12} & 0 \\
-\sin\theta_{12} & \cos\theta_{12} & 0 \\
0 & 0 & 1 \\
\end{array}\right),
\end{equation}
where $U$ is defined by $U^T M_\nu U={\rm diag}(m_1, m_2, m_3)$.
Under this requirement, the mixing angles should satisfy  
$\sin\theta_{23}\simeq 1/\sqrt 2$ and $\sin\theta_{12}\simeq 1/\sqrt 3$.
We fix these values just as $1/\sqrt 2$ and $1/\sqrt 3$, for simplicity. 
Since $M_\nu$ has only two non-zero mass eigenvalues,
we have two possibilities for the mass eigenvalues $m_{1,2,3}$ under
this setting: that is,   
$m_1=0$, $m_2=\sqrt{\Delta m_{\rm sol}^2}$ and $m_3=\sqrt{\Delta m_{\rm atm}^2}$
for normal hierarchy, and $m_3=0$,
$m_1=\sqrt{\Delta m_{\rm sol}^2}$ and $m_2=\sqrt{\Delta m_{\rm atm}^2}$
for inverse hierarchy.

We search solutions in which 5 parameters are consistently fixed by
applying the neutrino oscillation data to 
$\Delta m_{\rm sol}^2$ and $\Delta m_{\rm atm}^2$
and varying the value of $\sin\theta_{13}$ within the range $0\le
\sin^22\theta_{13} < 0.19$ \cite{chooz}.
As a result of this study, we can find consistent solutions only in the normal
hierarchy case. Inverse hierarchy seems not to be favored in this model.
In Fig.~2, we show an example of the solution with $\sin\theta_{13}=0.13$. 
Required values for $\Delta m_{\rm sol}^2$ 
and $\Delta m_{\rm atm}^2$ are plotted as regions sandwiched by two blue and 
red dashed lines, respectively. 
In the same figure, eigenvalues $m_2$ and $m_3$ are plotted by blue and 
red solid lines as functions of $\Lambda_1$.
The figure shows that each solid line crosses the required regions
for $\Delta m_{\rm sol}^2$ and $\Delta m_{\rm atm}^2$ at the same $\Lambda_1$.
This means that the model can have the consistent parameter sets for the
explanation of the neutrino oscillation data.
The fixed values of the 5 parameters
in $M_\nu$ are determined for this solution. They are listed in Table~2.
In the same table, as examples of other typical solutions, 
we also give the values of these parameters 
for the different $\sin\theta_{13}$. 
From these examples, we find that our model can have solutions with 
very small values of $\sin\theta_{13}$ and also solutions 
with its present bound value. 
In the last column of this table, we also show the predicted
values of the effective mass 
$\langle m_{ee}\rangle=\left|\sum_iU_{ei}^2m_i\right|$
which is the measure of the neutrinoless double $\beta$ decay.
These values are two order of magnitude smaller than the present bound.
Thus, it seems to be difficult to find the signature of this model in the
next generation experiments prepared for the neutrinoless double 
$\beta$ decay. 
 
\begin{figure}[t]
\begin{center}
\begin{tabular}{|l|l|l|l|l|l|l|}\hline
\multicolumn{1}{|c|}{$\sin\theta_{13}$} & \multicolumn{1}{|c|}{$\Lambda_1$} 
&\multicolumn{1}{|c|}{$\Lambda_2$} & \multicolumn{1}{|c|}{$y_1$} & 
\multicolumn{1}{|c|}{$y_2$} &\multicolumn{1}{|c|}{$\bar f$}
&\multicolumn{1}{|c|}{$\langle m_{ee}\rangle$} \\ \hline\hline 
0.13 & 0.028 & 0.371262 & 0.046855 & 0.767892 & 0.59699 &
$ 3.8\times10^{-3}$ \\ \hline
0.0205 & 0.023 & $-0.59241$ & 0.165318 & 1.2649 & 0.514604 &
$3.0\times 10^{-3}$ \\ \hline
0.2306 & 0.019 & $-0.747848$ & 0.580237 & 1.35967 & 0.681424 &
$1.5\times 10^{-5}$ \\ \hline
\end{tabular}
\end{center}
{\footnotesize Table 2~~Numerical values of the parameters for the solutions 
in the normal hierarchy case. All dimensional parameters are given by
the eV unit.}
\end{figure}  

\subsection{Other phenomenological features}
The model has various features discriminating from the MSSM at TeV regions.
Some of them are caused by the existence of extra color triplets and
extra Higgs doublets. The former makes it possible to generate two types 
of $\mu$-terms radiatively through the RGE effects on the soft masses of
the singlet scalars. They mix with the down type quarks as mentioned
before and we may find these extra color triplets in the 
LHC experiments \cite{extrag1, extrag2}.
The latter brings special structure for the charged lepton mass matrix and
also the extended chargino and neutralino sector through two types of 
$\mu$-term. We can expect that these features make it possible 
to distinguish the model from the MSSM. They may be
detected in future experiments. 

One-loop diagrams similar to the one for the neutrino masses in Fig.~1 
might cause additional contributes to other phenomena, which do not 
appear in the MSSM.
In fact, the lepton flavor violating processes such as 
$\mu\rightarrow e\gamma$ are known to give severe constraints on 
the original non-supersymmetric model for the radiative seesaw 
\cite{cdmmeg}.\footnote{A possibility 
to loose this tension in a non-SUSY framework is proposed in \cite{sty1,
sty2}.}   
Although the small neutrino masses are guaranteed by an 
extremely small Higgs coupling in the scalar potential 
in this original model,
the existence of the heavy doublet chiral superfield 
$H_{u3}$ makes the neutrino mass 
small in the present model.
Thus, we need no such a small coupling here.
This makes the nature of the new contributions to 
the lepton flavor violating processes very different in both cases.
Since these contributions are sufficiently suppressed due to the
heaviness of $H_{u_3}$ in the present model, the dominant constraints 
from the lepton flavor violating processes appear as the conditions for 
the ordinary contributions caused by the supersymmetry breaking sector. 
Thus, the neutrino mass generation is not affected by the FCNC
constraints. This is different from the original radiative seesaw model.
Because of the same reason, new contributions to the muon anomalous 
magnetic moment $\delta a_\mu$ due to the similar one-loop diagrams are 
also negligible. If we try to explain the presently reported experimental
value $\delta a_\mu=(30.2\pm 8.7)\times 10^{-10}$ \cite{expg2,mg2},
we need to find its origin in the contributions expected also in other
supersymmetric models.

Finally, we comment on a dark matter candidate.
Our model has the exact $Z_2$ symmetry, which can be identified with 
the $R$ parity. Since the even parity of this $Z_2$ symmetry are assigned to the
SM contents, dark matter is consider to be 
the lightest neutral field with the odd parity. 
Thus, the candidate is expected to be the lightest neutralino as
mentioned before.
In the present model, there are new neutralino components 
other than the MSSM ones. Since the lightest one is not related to the 
one-loop neutrino mass generation discussed before, 
its abundance has no direct relation with both the neutrino mass
generation and the constraint from the lepton flavor violating processes
such as $\mu\rightarrow e\gamma$.  
The nature of the lightest neutralino is determined by their mass matrix 
${\cal M}_{\cal N}$ given in Appendix B.  
Its detailed analysis is beyond the scope of
this paper. Here, we only point out an interesting possibility 
on this issue, especially, the relation between
the dark matter relic abundance and the PAMELA anomaly \cite{pamela}
which shows the positron excess at 10~GeV-100~GeV in the cosmic 
ray from the galactic center.

We consider a case in which a neutralino dominated by $\tilde N_1$ and 
$\tilde H_{u_1}$ or $\tilde N_2$ and $\tilde H_{u_2}$ is the lightest one. 
In that case their nature as the dark matter 
is completely different from that of the dark matter candidate in the MSSM.
Since such states have the Yukawa couplings 
$\kappa_1\tilde{\bar N}_1\tilde H_{u_1}H_{d_3}$, 
$\kappa_2\tilde{\bar N}_2\tilde H_{u_2}H_{d_3}$ and 
also U(1)$_\chi$ gauge interaction, 
they are expected to annihilate through these interactions 
into quarks and leptons.
Since $H_{d_3}$ couples only with leptons in the light fields as found
from $W_3$, the annihilation of the dark matter through 
the $s$-channel exchange of the neutral Higgs scalar $H_{d_3}^0$
can produce positrons but no antiprotons in the final states.\footnote{It
should be noted that $h^i_{\bar g}Q_i\bar g_3H_{d_3}$ can not contribute
to the dark matter annihilation since $\bar g_3$ is much heavier than the
supposed dark matter.} 
On the other hand, since quarks and leptons have non-zero U(1)$_\chi$ 
charge, the dark matter can annihilate to both quarks and leptons 
through this gauge interaction.
These aspects may allow us to understand the discrepancy between 
the values of the thermally averaged annihilation cross section 
$\langle\sigma v\rangle$ required to explain the relic abundance and the
PAMELA anomaly, respectively.
In order to explain this reason, we suppose that the model 
parameters can be arranged 
so that the U(1)$_\chi$ interaction is relevant only to the
determination of the relic abundance of the dark matter 
but the Yukawa interactions are only 
related to the PAMELA anomaly.
If the annihilation cross section due to the U(1)$_\chi$ interactions satisfies 
$\langle\sigma v\rangle\sim 3\times 10^{-26}$~cm$^3$/sec at the
freeze-out time of the dark matter, it is known to be suitable for the 
explanation of the relic abundance but smaller than the required one 
for the PAMELA anomaly by two or three order of magnitude. 
However, if the dark matter mass is almost equal to half of the mass 
eigenvalue of the neutral Higgs state dominated by $H_{d_3}^0$, 
the Breit-Wigner enhancement may make this annihilation cross section 
much larger in the present Galaxy \cite{bw1, bw2}. 
In that case, the final states of the annihilation is 
mainly composed of leptons.
This may allow the model to give a consistent explanation for this huge boost 
factor problem. 
Since the dark matter can not be so heavy in the present model as found
from the neutralino mass matrix (\ref{neutram}),
we need origins other than the dark matter annihilation for 
the explanation of the excess of positron and electron flux 
at higher energy regions 
observed in the Fermi-LAT experiment \cite{fermi}. 
We would like to discuss this issue quantitatively elsewhere.  

\section{Summary}
Present experimental data on dark matter and neutrino masses impose us
to extend the SM. In this paper we have considered a new possibility 
of such extensions in the framework of a supersymmetric model with an
extra U(1) symmetry at TeV regions, which is constructed on the basis 
of $E_6$. 
In the ordinary $E_6$ framework, unless the fields of higher
dimensional representations are introduced, 
only a unique example of the extra U(1) 
has been known to be consistent with large right-handed Majorana 
neutrino masses. In that case the small neutrino mass generation
can be considered on the basis of the seesaw mechanism.
In other types of low energy extra U(1) symmetry in the $E_6$ framework, 
however, 
it is difficult to make the right-handed neutrinos heavy enough keeping
the consistency with this TeV scale U(1) symmetry. Thus, the neutrino masses 
can not be small enough naturally as long as we follow the 
usual field assignment.
Although the extra U(1) symmetry in $O(1)$~TeV regions gives 
an elegant solution for the $\mu$ problem in the MSSM,
this solution can not be consistent with the small neutrino mass 
generation based on the seesaw mechanism except for the unique case
mentioned above.  

In this paper we have proposed a scenario in which a new type 
of extra U(1) symmetry in $E_6$ may give a consistent explanation for 
both the small neutrino mass generation and the $\mu$ problem.
By embedding the MSSM fields in the fundamental representation 
${\bf 27}$ in the different way among the generations, we have 
shown that both of them can be consistently explained. 
The small neutrino masses are generated by both the seesaw mechanism 
due to a heavy SU(2) doublet fermion and also one-loop effects 
similar to the radiative seesaw in the non-supersymmetric model. 
Since neutrino mass matrix has the constrained texture with the restricted
number of parameters, the model is predictive in the neutrino 
sector. Our numerical study shows that the model can be
consistent with all neutrino oscillation data. 
For that parameter set $\sin\theta_{13}$ can be predicted.
We have given such examples.
The model has other interesting phenomenological features, that is, 
the gauge coupling unification expected at a GUT scale, 
the existence of several extra fields which may be detected 
in the LHC experiment and others, and also the dark matter candidate which
may have different features from the one of the MSSM. 
These aspects seems to make the model interesting and 
also deserve further study.  
      
\vspace*{5mm}
\noindent
%{\Large Acknowledgement}\\
This work is partially supported by a Grant-in-Aid for Scientific
Research (C) from Japan Society for Promotion of Science (No.21540262).

\newpage
\noindent
{\Large\bf Appendix A}\\

\noindent
The charged lepton mass matrix in eq.~(\ref{qlmass}) is composed of Yukawa
couplings with two Higgs doublets. However, if Yukawa coupling constants
are assumed to satisfy simple relations, each mass eigenstate couples with
only one Higgs doublet as shown in eq.~(\ref{diag}). 
This guarantees to bring no additional origin for the lepton flavor violating 
processes. As such conditions, we may adopt
\begin{equation}
 \sum_{\beta=1}^3h_E^{\alpha\beta}h_E^{3\beta}=0 \qquad (\alpha=1,2).
\end{equation}
In this case, we can easily find that eq.~({\ref{diag}}) is realized, 
if we take a new basis for the right-handed charged leptons such as 
$\bar E^\prime=V\bar E$ where $V$ is defined as
\begin{equation}
V=\left(\begin{array}{ccc}
\frac{h_E^{32}}{\xi_1} & -\frac{h_E^{31}}{\xi_1} & 0\\
\frac{h_E^{31}h_E^{33}}{\xi_1\xi_2} &
 \frac{h_E^{32}h_E^{33}}{\xi_1\xi_2} &-\frac{\xi_1}{\xi_2} \\
\frac{h_E^{31}}{\xi_2} & \frac{h_E^{32}}{\xi_2} &\frac{h_E^{33}}{\xi_2} \\
\end{array}
\right),
\end{equation}
and $\displaystyle \xi_n=(
\sum_{\alpha=1}^{n+1}(h_E^{3\alpha})^2)^{1/2}$.
Yukawa coupling constants in this new basis are expressed as
\begin{eqnarray}
&&\tilde h_E^{11}=(h_E^{11}h_E^{32}-h_E^{12}h_E^{31})/\xi_1, \qquad
\tilde h_E^{12}=-h_E^{13}\xi_2/\xi_1, \nonumber \\
&&\tilde h_E^{21}=(h_E^{21}h_E^{32}-h_E^{22}h_E^{31})/\xi_1, \qquad
        \tilde h_E^{22}=-h_E^{23}\xi_2/\xi_1, \qquad
\tilde h_E^{33}=\xi_2.
\end{eqnarray}

\vspace*{5mm}

\noindent
{\Large\bf Appendix B}\\

\noindent
In this appendix we address both masses and mixings of the fields which
play important roles in the neutrino mass generation shown by Fig.~1.
They are expressed by $m_F$, $m_B$ and $\Lambda_B$ in the formulas for
$\Lambda_{2}$. Here we omit the suffices $a$ and $b$ which are written
in the text. In Fig.~1 $m_F$ and $\Lambda_B$ are drawn by the bulbs.

There are heavy colorless chiral superfields, which obtain masses through the 
couplings with ${\cal A}_7$ and $\bar{\cal A}_7$ as discussed in the text. 
Their fermionic components are $Z_2$ even as the ordinary quarks and leptons.
The effective superpotential for these heavy chiral superfields at the
weak scales are found to be given by
\begin{equation}
W_H=\gamma_1\phi H_{u_3}H_{d_1} 
+\gamma_2v_2 H_{d_1}\bar N_3 +\frac{1}{2}M_N\bar N_3^2, 
\end{equation}
where we list up dominant terms only and $M_N=O(\phi^2/M_{\rm pl})$.
The first two terms come from the last line of
the superpotential $W_2$. A mass term for $\bar N_3$ is induced through the
interaction in eq.~(\ref{majmass}).
Thus, the mass matrix for the neutral fermionic components of 
$H_{u_3}$, $H_{d_1}$, $\bar N_3$ can be expressed on the 
$(\tilde H_{u_3}^0,\tilde H_{d_1}^0,\tilde{\bar N}_3)$ basis as
\begin{equation}
{\cal M}_H=\left(\begin{array}{ccc}
0 & \gamma_1\phi & 0 \\
\gamma_1\phi & 0 & \gamma_2v_2 \\
0 & \gamma_2v_2 & M_N \\
\end{array}\right).
\label{h3}
\end{equation}

As found from the couplings of $\tilde H_{u_3}^0$ with $\nu_i$ in the
superpotential $W_3$, 
$\tilde H_{u_3}^0$ plays the similar role in the neutrino mass 
generation to the one of the right-handed neutrinos in the ordinary seesaw 
mechanism. It also contribute to the one-loop diagram for the neutrino mass
generation in Fig.~1. If we define a mixing matrix by 
$(V_H)^T{\cal M}_HV_H={\cal M}_H^{\rm diag}$,
we find that the mass eigenvalues and the mixing matrix $V_H$ 
are estimated as
\begin{equation}
{\cal M}_H^{\rm diag}\simeq{\rm diag}(\gamma_1\phi,~\gamma_1\phi,~M_N), \qquad
V_H\simeq\left(\begin{array}{ccc}
\frac{1}{\sqrt 2} & \frac{1}{\sqrt 2} & 0 \\
\frac{1}{\sqrt 2}e^{i\frac{\pi}{2}} & 
\frac{1}{\sqrt 2}e^{i\frac{\pi}{2}} & 0 \\   0 & 0 & 1 \\
\end{array}\right),  
\end{equation}
where we use the relation $\phi, ~M_N \gg v_2$ to derive these results. 
If we express $\Lambda_2$ in
eq.~(\ref{scale}) as $\Lambda_2=f(m_F,m_B)$, $\Lambda_2$ can be written
by using the mass eigenstates derived above as
\begin{equation}
\Lambda_2=\sum_{a=1}^3\left\{(V_H)_{1a}\right\}^2f(M_a,m_B)=f(\gamma_1\phi,m_B).
\end{equation}
Thus, the effective mass $m_F$ of the internal fermion in Fig.~1(a) can be 
estimated as $m_F\sim \gamma_1\phi$.

Next, we represent scalar partners of these chiral superfields as 
$\Phi=(H_{u_3}^0, H_{d_1}^0,\bar{N_3})$ and 
define their mass terms by
\begin{equation}
-{\cal L}_{\Phi}=\frac{1}{2}\left(\Phi^\dagger {\cal M}_B^2
\Phi + \Phi^T {\cal M}_m^2\Phi\right)+ {\rm h.c.}, 
\label{smass}
\end{equation}
where these mass matrices can be expressed as
\begin{eqnarray}
&&{\cal M}_B^2=\left(\begin{array}{ccc}
\gamma_1^2\phi^2+m_s^2 & 0 & \gamma_1\gamma_2\phi v_2  \\
0 & \gamma_1^2\phi^2+\gamma_2^2v_2^2 + m_s^2 &  \gamma_2v_2M_N \\
\gamma_1\gamma_2\phi v_2 & \gamma_2v_2M_N & 
M_N^2 +\gamma_2^2v_2^2 +m_s^2 \\
\end{array}\right), \nonumber \\
&&{\cal M}_m^2=\left(\begin{array}{ccc}
0 &  A\gamma_1 \phi & 0 \\
A\gamma_1 \phi & 0 & A\gamma_2 v_{2} \\
0 & A\gamma_2 v_{2} & A M_N \\
\end{array}\right).
\label{bmass}
\end{eqnarray}
In these mass matrices we introduce the supersymmetry 
breaking universal soft scalar masses
$m_s^2$ and also the universal soft supersymmetry 
breaking parameter $A$ for the scalar 
trilinear couplings. Since $\phi$ constitutes a $D$-flat direction for
U(1)$_\psi$, there are no $D$-term contributions to the scalar masses.
By using these results, the effective mass $m_B$ and the mixing $\Lambda_B$ 
appeared in the internal scalar line of Fig.~1(b) are estimated as
\begin{equation}
m_B^2\simeq (\gamma_1\phi)^2, \qquad 
\Lambda_B^2\simeq\frac{A^5(\gamma_2v_2)^2}{M_N^3(\gamma_1\phi)^2},
\label{fmass} 
\end{equation}
where we again take account of $\phi \gg v_2, m_s$ to derive these results.

The model contains a lot of $Z_2$ odd neutral fermions and their 
$Z_2$ even scalar partners.
One-loop diagrams for the neutrino mass generation in Fig.~1 include 
chiral superfield $\bar N_2$ which is a member of them.
Thus, the mixings of its fermionic partner with other $Z_2$ odd fermions 
are crucial for the estimation of this diagram.
These light $Z_2$ odd neutral fermions include the SU(2) gaugino $\tilde W_3$,
the U(1)$_Y$ gaugino $\tilde B$, the U(1)$_\chi$ gaugino $\tilde\lambda_\chi$
and several fermionic components of chiral superfields. 
Relevant terms in the superpotential $W_3$ and $W_2^{NR}$ in 
eq.~(\ref{nr2}) are
\begin{equation}
W_{\cal N}=\lambda_1 S_2H_{u_2}H_{d_2}+\lambda_2S_3H_{u_2}H_{d_3}
+\kappa_1\bar N_1H_{u_1}H_{d_3}+\kappa_2 \bar N_2H_{u_2}H_{d_3}
+\kappa_3\epsilon^2\phi H_{u_1}H_{d_2},
\end{equation}
where we introduce a new coupling constant $\kappa_3$.
If we take a basis for these neutral fermions as
$$\tilde{\cal N}=(\tilde W_3, \tilde B, \tilde\lambda_\chi,\tilde H_{u_2}, 
\tilde H_{d_2}, \tilde H_{d_3}, \tilde S_2, \tilde S_3, 
\tilde H_{u_1}, \tilde N_1, \tilde N_2),
$$ 
their tree-level mass matrix ${\cal M}_{\cal N}$ can be expressed as
\begin{equation}
\left(\begin{array}{ccccccccccc}
M_2 & 0 & 0 & -\frac{g_2v_2}{\sqrt{2}} & \frac{g_2v_{1a}}{\sqrt{2}}&
\frac{g_2v_{1b}}{\sqrt{2}} & 0 & 0 & 0 & 0 & 0 \\ 
0 & M_1 & 0 & \frac{g_1v_2}{\sqrt{2}} & -\frac{g_1v_{1a}}{\sqrt{2}}&
-\frac{g_1v_{1b}}{\sqrt{2}} & 0 & 0 & 0 & 0 & 0 \\
0 & 0 & M_\xi & \frac{q_2g_\xi v_2}{\sqrt{2}} & 
\frac{q_1g_\xi v_{1a}}{\sqrt{2}}&
\frac{q_1g_\xi v_{1b}}{\sqrt{2}} & 
\frac{q_Sg_\xi u}{\sqrt{2}} & \frac{q_Sg_\xi u^\prime}{\sqrt{2}} 
& 0 & 0 & 0 \\
-\frac{g_2v_2}{\sqrt{2}} &\frac{g_1v_2}{\sqrt{2}} & 
\frac{q_2g_\xi v_2}{\sqrt{2}} & 0 & \mu & \mu^\prime & 
\lambda_1 v_{1a} & \lambda_2 v_{1b} & 0 & 0 & \kappa_2v_{1b} \\
\frac{g_2v_{1a}}{\sqrt{2}} & -\frac{g_1v_{1a}}{\sqrt{2}} & 
 \frac{q_1g_\xi v_{1a}}{\sqrt{2}} & \mu  & 0 & 0 
& \lambda_1 v_2 & 0 & \kappa_3\epsilon^2\phi & 0 & 0 \\
\frac{g_2v_{1b}}{\sqrt{2}} & -\frac{g_1v_{1b}}{\sqrt{2}} & 
 \frac{q_1g_\xi v_{1b}}{\sqrt{2}} & \mu^\prime & 0 &0 & 0 & \lambda_2 v_2 
& 0  & 0 & \kappa_2v_2 \\
0 & 0 &  \frac{q_Sg_\xi u}{\sqrt{2}} & \lambda_1v_{1a} & \lambda_1 v_2 
& 0  & 0 & 0 & 0  & 0 & 0 \\
0 & 0 &  \frac{q_Sg_\xi u^\prime}{\sqrt{2}} & \lambda_2v_{1b} & 0 & 
\lambda_2 v_2 & 0  & 0 & 0 & 0 & 0 \\
0 & 0 & 0 & 0 & \kappa_3\epsilon^2\phi & 0  & 0  & 0 & 0 & 
\kappa_1v_{1b} & 0 \\
0 & 0 & 0 & 0 & 0 & 0  & 0  & 0 & \kappa_1v_{1b} & 0 & 0 \\
0 & 0 & 0 & \kappa_2v_{1b} & 0 & \kappa_2v_2  & 0  & 0 & 0 & 0 & 0   \\
\end{array}\right).
\label{neutram}
\end{equation}
Each neutralino component $\tilde{\cal N}_n$ is related to the mass eigenstates 
$\tilde\chi_a$ through
\begin{equation}
\tilde{\cal N}_n=\sum_a (V_N)_{na}\tilde\chi_a,
\end{equation}
where $V_N$ is the mixing matrix which diagonalizes the neutralino mass matrix
${\cal M}_N$ defined in eq.~(\ref{neutram}) as 
$V_N^T{\cal M}_NV_N={\rm diag}(\tilde M_1,\tilde M_2, \cdots, \tilde M_{11})$.
In this neutralino case, $\Lambda_2$ can be expressed as
\begin{equation}
\Lambda_2=\sum_{a=1}^{11}\left\{(V_N)_{9a}\right\}^2 f(\tilde M_a,m_B). 
\end{equation}
Since details are dependent on a lot of parameters,
we can not give analytic expression for the effective mass $m_F$.
However, it is obvious that $m_F$ takes a weak scale value $m_w$.
This rough estimation is enough for the present purpose.

In order to estimate $m_B$ and $\Lambda_B$ for the scalar partners of
these fermions, we need to take account of the 
couplings $\lambda_2S_3H_{u_2}H_{d_3}$ and
$\kappa_2\bar N_2H_{u_2}H_{d_3}$ in $W_{\cal N}$.
If we adopt a basis as $\Phi=(H_{u_2}^0, H_{d_3}^0, S_3, \bar
N_2)$ and express the mass terms by eq.~(\ref{smass}), 
${\cal M}_B^2$ and ${\cal M}_m^2$ in this case 
can be written as
\begin{eqnarray}
&&{\cal M}_B^2=\left(\begin{array}{cccc}
\mu^{\prime 2}+\bar\lambda_2^2v_{1b}^2+m_s^2 & 0 & 0 & 
2\kappa_2\mu^\prime v_2 \\
0 & \mu^{\prime 2}+\bar\lambda_2^2v_2^2+m_s^2 & 0 &
2\kappa_2\mu^\prime v_{1b} \\
0 & 0 & \lambda_2^2\bar v_b^2+m_s^2 
& \lambda_2\kappa_2\bar v_b^2 \\
2\kappa_2\mu^\prime v_2 & 2\kappa_2\mu^\prime v_{1b} & 
\lambda_2\kappa_2\bar v_b^2 & \kappa_2^2\bar v_b^2+m_s^2 \\
\end{array} \right), \nonumber \\
&&{\cal M}_m^2=\left(\begin{array}{cccc}
0 & A\mu^\prime & A\lambda_2v_{1b} & A\kappa_2v_{1b} \\
A\mu^\prime & 0 & A\lambda_2v_2 & A\kappa_2v_2 \\
A\lambda_2v_{1b} & A\lambda_2v_2 & 0 & 0 \\
A\kappa_2v_{1b} & A\kappa_2v_2 & 0 & 0 \\ \end{array}\right),
\label{ah3}
\end{eqnarray}
where $\bar v_b^2=v_2^2+v_{1b}^2$ and $\bar\lambda^2=\lambda_2^2+\kappa_2^2$.
Although there are D-term contributions to ${\cal M}_B^2$,
they are not written explicitly here.
From these mass matrices, the effective mass $m_B^2$ and the mixing 
$\Lambda_B^2$ in Fig.~1(a) are approximately estimated as
\begin{equation}
m_B^2\simeq m_s^2, \qquad 
\Lambda_B^2=
\frac{\kappa_2^2A^3\mu^\prime v_2v_{1b}}{m_s^4}. 
\end{equation}
We find that both of these take values of order of the weak scale.

\end{document}